\documentclass[sigconf]{acmart}

\usepackage{soul}
\usepackage{xcolor}
\usepackage{xspace}
\usepackage{url}
\usepackage{graphicx}
\usepackage{listings,multicol}
\usepackage[caption=false]{subfig}  
\usepackage[export]{adjustbox}
\usepackage{amsmath}

\usepackage{amssymb}
\usepackage{textcomp}
\usepackage{breakurl} 
\usepackage{listings}
\usepackage{hyperref}
\usepackage{cleveref}
\usepackage{algorithm}
\usepackage[noend]{algpseudocode}
\usepackage{dblfloatfix} 
\usepackage{enumitem}
\usepackage{kantlipsum}
\usepackage{multirow}
\usepackage{array}
\usepackage{booktabs}
\usepackage{tablefootnote}
\usepackage{color}
\usepackage{colortbl}

\usepackage{balance}
\usepackage{tikz}

\newcommand*\circled[1]{\tikz[baseline=(char.base)]{
            \node[shape=circle,draw,inner sep=2pt] (char) {#1};}}
\crefformat{footnote}{#2\footnotemark[#1]#3}

\graphicspath{ {./img/} }

\newcommand{\toolName}{\textit{KOALA}\xspace}

\title[KOALA: a Configurable Tool for Collecting IDE Data When Solving Programming Tasks]{KOALA: a Configurable Tool for Collecting\\IDE Data When Solving Programming Tasks}

\ccsdesc[500]{Social and professional topics~Computing education}
\ccsdesc[300]{Human-centered computing~Interactive systems and tools}

\AtBeginDocument{%
  }

\setcopyright{acmlicensed}
\copyrightyear{2018}
\acmYear{2018}
\acmDOI{XXXXXXX.XXXXXXX}

\acmConference[Conference acronym 'XX]{Make sure to enter the correct
  conference title from your rights confirmation emai}{June 03--05,
  2018}{Woodstock, NY}
\acmISBN{978-1-4503-XXXX-X/18/06}

\begin{document}

\author{Daniil Karol}
\affiliation{%
  \institution{JetBrains Research}
  \city{Berlin}
  \country{Germany}}
\email{daniil.karol@jetbrains.com}

\author{Elizaveta Artser}
\affiliation{
  \institution{\textit{JetBrains Research}}
  \city{Munich}
  \country{Germany}
}
\email{elizaveta.artser@jetbrains.com}

\author{Ilya Vlasov}
\affiliation{%
  \institution{\textit{JetBrains Research}}
  \city{Belgrade}
  \country{Serbia}}
\email{ilya.vlasov@jetbrains.com}

\author{Yaroslav Golubev}
\affiliation{%
  \institution{JetBrains Research}
  \city{Belgrade}
  \country{Serbia}}
\email{yaroslav.golubev@jetbrains.com}

\author{Hieke Keuning}
\affiliation{%
  \institution{Utrecht University}
  \city{Utrecht}
  \country{The Netherlands}}
\email{h.w.keuning@uu.nl}

\author{Anastasiia Birillo}
\affiliation{%
  \institution{JetBrains Research}
  \city{Belgrade}
  \country{Serbia}}
\email{anastasia.birillo@jetbrains.com}

\begin{abstract}

Collecting data of students solving programming tasks is incredibly valuable for researchers and educators. It allows verifying that the students correctly apply the features and concepts they are taught, or finding students’ misconceptions. However, existing data collection tools have limitations, \textit{e.g.}, no control over the granularity of the collected code, not collecting the specific events of the programming environment used, and overall being hard to configure. 

To overcome these limitations, we propose \toolName, a convenient and highly configurable tool for collecting code snapshots and feature usage from students solving programming tasks in JetBrains IDEs. The plugin can be installed in IDEs and configured to provide the students with the necessary tasks, enable or disable certain IDE features like code completion, and run surveys. During problem solving, the plugin collects code snapshots at the configured granularity, all IDE actions like running and debugging, as well as some data not collected in prior works, like employed hotkeys and switching focus between files. The collected data is sent to the server that comes with the tool, where it is stored and can be converted to the standardized ProgSnap2 format. To showcase the tool, we collected data from 28 students solving tasks in two courses within the IDE, highlighting some insights from this data.

\end{abstract}

\keywords{data collection, in-IDE learning, code snapshots, activity data}

\maketitle

\begin{table*}[h!]
\centering
\caption{Comparison of existing tracking tools with \toolName. Data marked with (*) requires installing an external plugin.}
\begin{tabular}{@{}lllllll@{}}
\toprule
\multicolumn{1}{c}{\multirow{2}{*}{\textbf{Tool}}}                   & \multicolumn{1}{c}{\multirow{2}{*}{\textbf{Target IDE}}}          & \multicolumn{2}{c}{\textbf{Tracking Data}} & \multicolumn{1}{c}{\multirow{2}{*}{\textbf{Granularity}}} & \multicolumn{1}{c}{\multirow{2}{*}{\textbf{Customization}}} & \multicolumn{1}{c}{\multirow{2}{*}{\textbf{State}}} \\\cmidrule(lr){3-4}
\multicolumn{1}{c}{} & \multicolumn{1}{c}{} & \multicolumn{1}{c}{\textbf{Code}} & \multicolumn{1}{c}{\textbf{Events}} & \multicolumn{1}{c}{} & & \\ \midrule

\rowcolor{gray!20}\begin{tabular}[c]{@{}l@{}}\textbf{Dev Event}\\ \textbf{Tracker}~\cite{devEventTracker}\end{tabular} & Eclipse & \begin{tabular}[c]{@{}l@{}}Code \\ snapshot\end{tabular}  & Run/Debug & \begin{tabular}[c]{@{}l@{}}Each \\ program \\ save\end{tabular} & None & \begin{tabular}[c]{@{}l@{}}Last update 6 years\\  ago~\cite{det_github}\end{tabular}    \\

\begin{tabular}[c]{@{}l@{}}\textbf{Test My}\\ \textbf{Code}~\cite{testMyCode}\end{tabular} & \begin{tabular}[c]{@{}l@{}}NetBeans\\JetBrains\\ IDEs\end{tabular} & \begin{tabular}[c]{@{}l@{}}Code\\ snapshot\end{tabular} & \begin{tabular}[c]{@{}l@{}}- Interaction with UI\\ - Run/Debug\end{tabular} & \begin{tabular}[c]{@{}l@{}}Each\\ keystroke\end{tabular} & None & 
\begin{tabular}[c]{@{}l@{}} - The server side is\\being actively supported\\for testing solutions~\cite{test_my_code_server_github}\\ - IDEs plugins were updated\\5 years ago~\cite{test_my_code_netbeans_github, java_fx}\end{tabular} \\

\rowcolor{gray!20}\begin{tabular}[c]{@{}l@{}}\textbf{Task}\\ \textbf{Tracker}~\cite{lyulina2021tasktracker}\end{tabular}
& \begin{tabular}[c]{@{}l@{}}JetBrains\\ IDEs\end{tabular} & \begin{tabular}[c]{@{}l@{}}Code \\ snapshot\end{tabular} & \begin{tabular}[c]{@{}l@{}}- Run/Debug*\\ - Keyboard events*\\ - Mouse events*\\ - IDE actions*\end{tabular} & \begin{tabular}[c]{@{}l@{}}Each \\ keystroke\end{tabular} & \begin{tabular}[c]{@{}l@{}}Change task text in\\a predefined template\end{tabular}  & \begin{tabular}[c]{@{}l@{}}- Last update 4 years\\  ago~\cite{tasktracker_github}\\ - Two extra plugins~\cite{ActivityTracker, java_fx} \\ needed to track data\end{tabular} \\

\midrule

\textbf{\toolName}                                                   & \begin{tabular}[c]{@{}l@{}}JetBrains\\ IDEs\end{tabular} & \begin{tabular}[c]{@{}l@{}}Step-by-step\\code \\ snapshot\end{tabular}              & \begin{tabular}[c]{@{}l@{}}- Interaction with UI\\ - Run/Debug\\ - Keyboard events\\ - Mouse events\\ - IDE actions\\ - Hotkeys\end{tabular} & \begin{tabular}[c]{@{}l@{}}Each \\ keystroke\end{tabular}          & \begin{tabular}[c]{@{}l@{}}- Files to track\\ - IDE Settings\\ - Granularity\\ - Custom Surveys\\ - Events frequency\end{tabular} & \begin{tabular}[c]{@{}l@{}}Supports the latest versions \\  of all JetBrains IDEs \end{tabular}              \\ \bottomrule
\end{tabular}
\label{table:motivation:background}
\end{table*}

\section{Introduction}
\label{section:introduction}

The use of collecting programming data through development tools has grown significantly~\cite{edwards2023review, cao2020tool}, serving various purposes such as tool improvement~\cite{birillo2024one}, analyzing user interaction~\cite{interaction}, and enhancing programming education. In education, tracking students' interactions with IDEs like BlueJ~\cite{BlueJ} or NetBeans~\cite{NetBeans}  helps to understand task-solving approaches and tool usage. This involves logging user actions (\textit{e.g.}, clicks~\cite{devEventTracker}), keystrokes~\cite{edwards2023review}, or code changes~\cite{testMyCode, devEventTracker, BlackBox}. Such data aids in analyzing learning challenges~\cite{blackboxMisconceptions} and readiness for real-world tasks~\cite{RealWork}.  

Many tools exist for collecting data from students solving programming tasks~\cite{testMyCode, devEventTracker, lyulina2021tasktracker}, as well as extensive datasets of already collected data~\cite{BlackBox, datasetMOOC}. These tools are designed for popular IDEs, \textit{e.g.}, Eclipse, NetBeans, or BlueJ. However, they lack customization for educational scenarios, \textit{e.g.}, enabling or disabling code completion or style checks, as well as flexible control over the frequency and granularity of data collection, limiting researchers' ability to study student behavior in detail. 

To overcome the shortcomings of existing tools, we propose \toolName, a tool for collecting code snapshots and IDE activity data when solving programming tasks in JetBrains IDEs. \toolName is designed to be easy to use and adaptable for specific educational and research purposes by using configuration files, without changing the plugin's source code. The proposed tool supports all JetBrains IDEs, such as IntelliJ IDEA~\cite{intellijIdea}, PyCharm~\cite{pycharm}, CLion~\cite{cline}, which unifies the data collection process for programming tasks in different languages. 
We distribute this tool under the MIT license to provide as much flexibility to the community as possible.\footnote{\toolName tool source code: \url{https://github.com/JetBrains-Research/KOALA}}

To demonstrate the versatility and usefulness of \toolName, we collected a dataset with activities of 28 students solving programming tasks in two in-IDE courses~\cite{birillo2024bridging}. In total, the data consists of over 127 thousand code snippets and 585 thousand activity events. As a simple example of potential data analysis on the novel types of collected data, we highlight what refactoring hotkeys were the most popular among students. The dataset can be found in supplementary materials~\cite{supplementary}.

\toolName has also already shown its practical usefulness in one of our previous studies~\cite{birillo2024one}.
In that paper, we extended the functionality of one of the JetBrains IDEs to provide personalized help to students during learning. \toolName was used to collect all the information about the interaction with the IDE during the problem solving session in order to assess the quality of the proposed approach.

\newcommand{\customicon}[2]{%
    \includegraphics[width=0.08\textwidth]{#1} \raisebox{0.1\height}{#2}%
}

\section{Background}
\label{section:motivation}

\begin{figure*}[t]
    \centering
    \includegraphics[width=0.9\linewidth]{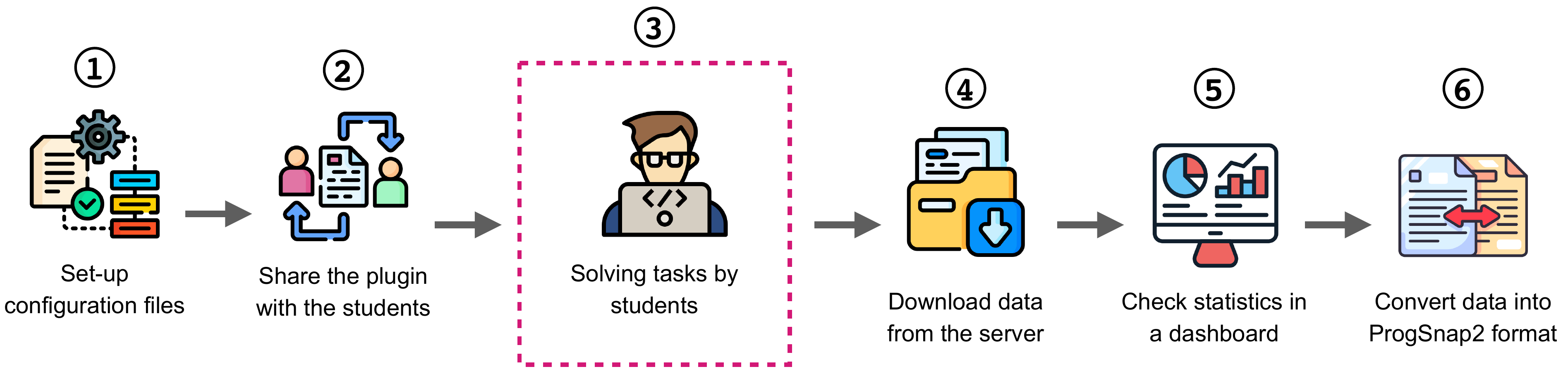}
     \vspace{-0.3cm}
    \caption{The general pipeline of the data gathering process with \toolName.}
     \vspace{-0.3cm}
    \label{fig:pipeline}
\end{figure*}

\subsection{Data Collection Tools}

Collecting data about user activities in development environments is an active research area that already produced a large number of studies and tools~\cite{ihantola2015educational, kiesler2023s,BlackBox,datasetMOOC}. The data can vary widely, ranging from which functions students use in the development environment~\cite{testMyCode, devEventTracker, BlackBox} to tracking the progress of students solving a specific task~\cite{datasetMOOC}.
Most existing tools are designed for use with Eclipse and NetBeans IDEs~\cite{testMyCodeAutomation}. However, they lack compatibility with the widely used JetBrains IDEs. This section provides a brief overview of the available tools for collecting code snapshots and IDE activity, with a particular focus on those developed for JetBrains IDEs. A summary of the reviewed tools is presented in \Cref{table:motivation:background}. 

\textbf{\textit{DevEvent Tracker}}~\cite{devEventTracker} is an Eclipse plugin designed to monitor and collect data on student progress by tracking editing, running, debugging events, errors, and refactorings. However, its granularity is limited, as snapshots are only captured upon saving, missing real-time changes. In addition, it lacks IDE customization options. 

\textbf{\textit{Test My Code}}~\cite{testMyCode} is a set of plugins for the NetBeans~\cite{NetBeans} and JetBrains IDEs that collect real-time programming data and are popular among researchers~\cite{testMyCodeAutomation, testMyCodeExploring}. 
Plugins track interactions with various IDE interface elements, as well as running and debugging programs. 
However, they lack customization of the IDE settings for a specific task, potentially limiting the researchers' flexibility.
Also, both plugins are not supported anymore~\cite{test_my_code_netbeans_github, test_my_code_ij_github} and only a server that checks students submissions by running tests is available~\cite{test_my_code_server_github}.

Several years ago, the \textbf{\textit{TaskTracker}} tool~\cite{lyulina2021tasktracker} was introduced for tracking code changes and activity events during the solving of programming tasks in JetBrains IDEs. It allows the tracking of all step-by-step changes in students' files together with actions that are performed inside the IDE, like debugging or copy-pasting events. The plugin and the collected dataset have been used in multiple studies~\cite{roest2024next, kasatskii2023effect, lohr2024let}, however, it has several limitations. 

First, the plugin installation process is complex, as it requires students to install both \textit{TaskTracker} and \textit{Activity Tracker}~\cite{ActivityTracker} plugins. \textit{Activity Tracker} plugin starts tracking only if the student turns it on manually. Additionally, students using IDE versions beyond the outdated 2020.2 version must install a third plugin~\cite{java_fx} to support the \textit{JavaFX platform}~\cite{java_fx_itself} for \textit{TaskTracker}'s UI. 

The second drawback of the plugin is its limited configuration options. The only configuration option is to update the content of the provided tasks (such as text or input and output data) and the data for student surveys. Any other potential customization of the data collection process---\textit{e.g.}, granularity, IDE settings, the order of tasks---requires changing the plugin's source code~\cite{kasatskii2023effect}, thus significantly decreasing the tool's usability.

The goal of the present work is to create a new open-source tool for JetBrains IDEs to gather detailed data about students' programming behaviors during problem solving, while overcoming the disadvantages of the tools described in this section. The tool should be flexible in configuration and installation, capable of collecting a wide range of data about the students' interaction with the IDE, and intuitively understandable for both researchers and students.

\subsection{JetBrains IDEs}
\label{section:jetbrains_ide}

JetBrains~\cite{jetbrains} is one of the most popular IDE providers for professional developers. JetBrains provides IDEs for different languages, such as IntelliJ IDEA~\cite{intellijIdea} for Java and Kotlin, PyCharm~\cite{pycharm} for Python, and CLion~\cite{cline} for C++. Despite the fact that all these IDEs are designed for professionals, they are also actively used for educational purposes---JetBrains provides free educational licenses for all IDEs to teachers and students upon request~\cite{jbEduLicences}. JetBrains has also recently introduced the in-IDE learning approach~\cite{birillo2024bridging} via the JetBrains Academy plugin~\cite{jetbrains-academy-plugin}, which allows the entire learning process to be moved into the professional environment. 
Finally, many researchers develop their own JetBrains IDE extensions to make the learning process easier for students~\cite{app-courses-plugin, orion-plugin}.
All of this points to the importance of having a data collection tool that would support all these IDEs in the educational environment.
\section{KOALA: Usage Pipeline}
\label{section:usage}

While \toolName is developed from scratch, its plugin-server architecture is inspired by that of \textit{TaskTracker}~\cite{lyulina2021tasktracker}, consisting of:
\begin{itemize}
    \item An IDE \textbf{plugin} for monitoring the progress and extracting the data within the IDE.
    \item A \textbf{server} that collects the data remotely, gathering solutions from multiple students and storing them in a database.
    \item A \textbf{dashboard} to facilitate the analysis of the collected data and provide basic statistics and insights.
    \item A \textbf{converter} to process the collected data and convert it into the \textit{ProgSnap2}~\cite{progsnap2} format.
\end{itemize}

The general pipeline of the data gathering process is presented in Figure~\ref{fig:pipeline}. Firstly, the researcher or the educator sets up the plugin using YAML configuration files (see \circled{1}), described in detail in Section~\ref{section:Mnemosyne:plugin}. These configurations include the contents of tasks, the IDE features that should be disabled, how often the code changes should be collected, etc. The configuration files allow to adapt the plugin for any experiment by specifying its order of events, description, and settings, without changing the plugin's source code. When this is done, the plugin is ready for usage and should be distributed among the students and installed into their IDEs (see \circled{2}). \toolName works with different IDEs, depending on the studied programming language, but the configuration files are the same for all of them. 

In the next step, the students are solving the tasks using the installed plugin (see \circled{3}). \toolName shows a tool window in the IDE that describes the task and provides a file to solve it in. It can also show additional information to the student or run surveys. The plugin only tracks files that were specified in the configuration files to avoid accidentally collecting any personal data from the students. After the tasks are solved, the students send the data to the server by clicking the ``Submit'' button, and all files are parsed and stored in a database on the remote server (see \circled{4}). Additionally, the files are saved locally to prevent loss of progress in case of unreliable internet connection. 

Next, the researcher or educator who designed the study could use a visualization dashboard to review the basic statistics about the collected data (see \circled{5}). The statistics include the total number of participants and general study information, as well as detailed insights into frequently used IDE actions or IDE windows and the time spent focusing on the particular task files during the study. This dashboard allows researchers to better interpret the data and draw preliminary insights.

Finally, in the last step, the collected data is converted into the \textit{ProgSnap2}~\cite{progsnap2} format (see \circled{6}). The main purpose of adding this step is to provide an opportunity to use existing data analysis tools for this format instead of writing them from scratch, thus simplifying \toolName's adoption by researchers and educators. With the data collected, it is now possible to move on to its analysis.

\begin{figure*}[t]
    \centering
    \includegraphics[width=\linewidth]{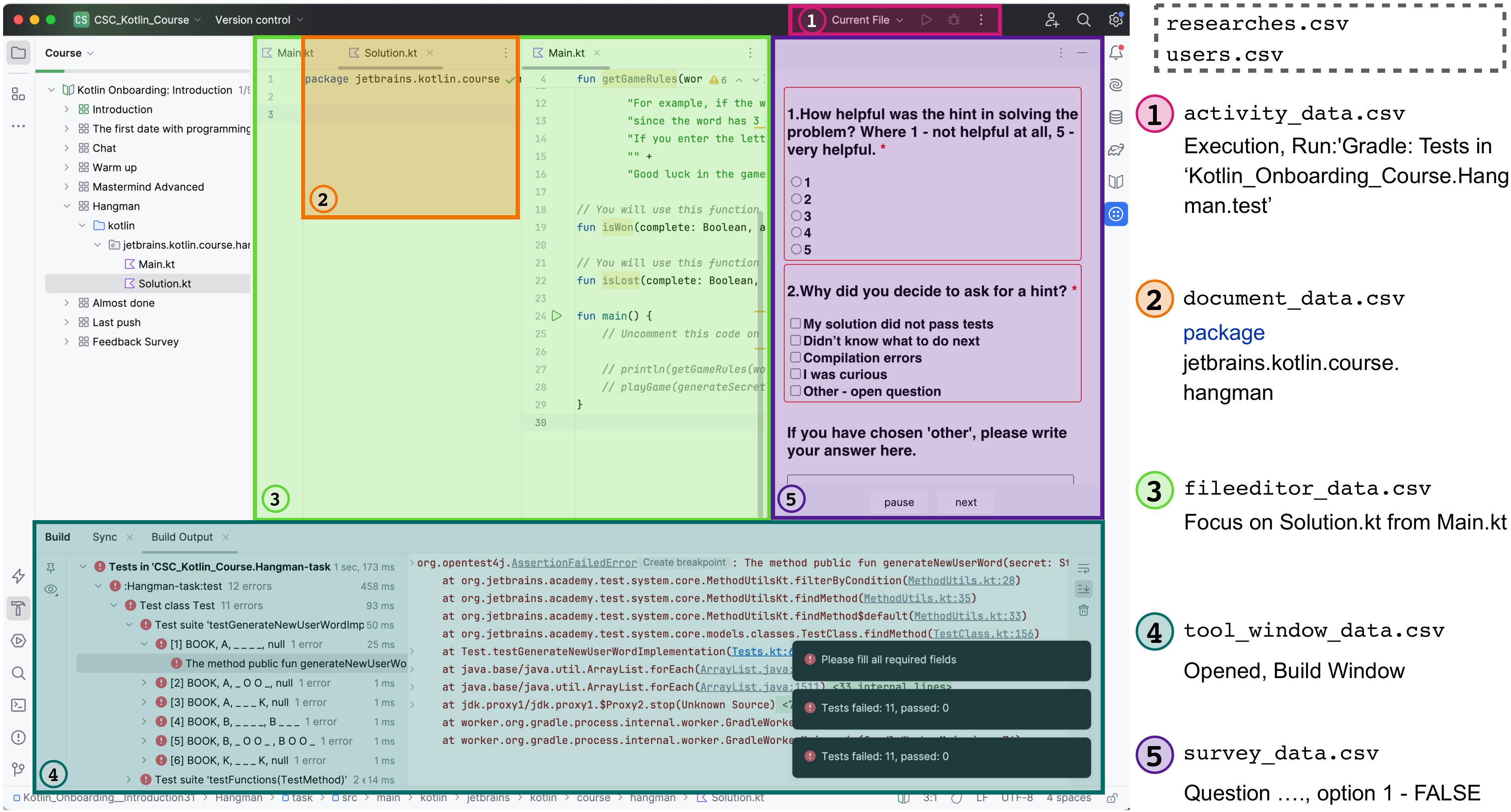}
    \vspace{-0.3cm}
    \caption{Examples of data collected using \toolName: (1) activities performed in the IDE, such as running, debugging, etc., (2) current student code, (3) opening, closing, and refocusing of files, (4) opened tool windows, and (5) survey data. On the right, you can see how different data is written to different CSV files, you can find the full dataset in supplementary materials~\cite{supplementary}.}
     \vspace{-0.3cm}
    \label{fig:collected_data}
\end{figure*}

\section{KOALA: Internal Design}
\label{section:Mnemosyne}

In this section, we will dive deeper into the details of how each component of \toolName is implemented.

\subsection{Plugin}\label{section:Mnemosyne:plugin}

The \toolName plugin is written for all JetBrains IDEs using the IntelliJ Platform Plugin SDK under the hood. This platform allows for the development of plugins for JetBrains IDEs~\cite{intellijSDK}, including the ability to interact with the IDE in various ways, such as adding tool windows for plugin interaction, handling different IDE events, and more. For the UI component of the plugin, we used a combination of Java Chromium Embedded Framework (JCEF)~\cite{jcef} and Swing~\cite{swing} technologies: JCEF was used to visually display all the educational information specified in the configuration files during the plugin setup, such as assignment descriptions and quizzes, and Swing was used to create interactive elements such as buttons to move on to the next step of the scenario.
Upon acceptance, the plugin will be fully open-sourced and available on GitHub.

\textbf{Collected data.} The plugin is capable of collecting a wide range of data concerning user interactions with the IDE. Some examples are shown in Figure~\ref{fig:collected_data}: \circled{1} activities performed in the IDE, \textit{e.g.}, the \textit{run} action; \circled{2} current student code; \circled{3} opening, closing, and refocusing of files; \circled{4} opened tool windows, \textit{e.g.}, the \textit{Build} window; \circled{5} survey data if a survey was set up. The plugin also saves some data that was not available in existing tools, such as the used hotkeys, \textit{i.e.}, keyboard shortcuts used to call various actions. The tracking of student interactions with the IDE is carried out with a certain granularity, which is configured separately for code snapshots and for various actions (see more details below).

The described data is collected in CSV files. The examples of individual data points are shown in Figure~\ref{fig:collected_data} on the right, and the full collected dataset can be found in supplementary materials~\cite{supplementary}. The data is sent to the server at times when a student decides to pause the task or completes it, allowing the plugin to operate under conditions of an unstable internet connection. Additionally, the plugin allows you to collect not only the previously described data, which is tracked by  \toolName itself, but also the data produced by other plugins or tools. For this purpose, in the configuration settings, it is possible to specify the location of the file that will be produced by a third-party tool, and then the file will be sent to the server and further processed. This allows \toolName to obtain an even richer view of students’ interactions with the IDE.

\textbf{Configuration}. To ensure that the plugin can be easily configured for different scenarios, it was decided to use YAML configuration files~\cite{yaml}. These files control various aspects of the plugin's behavior, such as the overall order of events, changing IDE settings during task completion, the granularity of tracking code changes, and others. In fact, these files control almost all aspects of the plugin's behavior, ensuring that \toolName can support more diverse experiments than previous tools. Let us now briefly describe the available configuration files.

\begin{figure}[t]
    \centering
    \includegraphics[width=\linewidth]{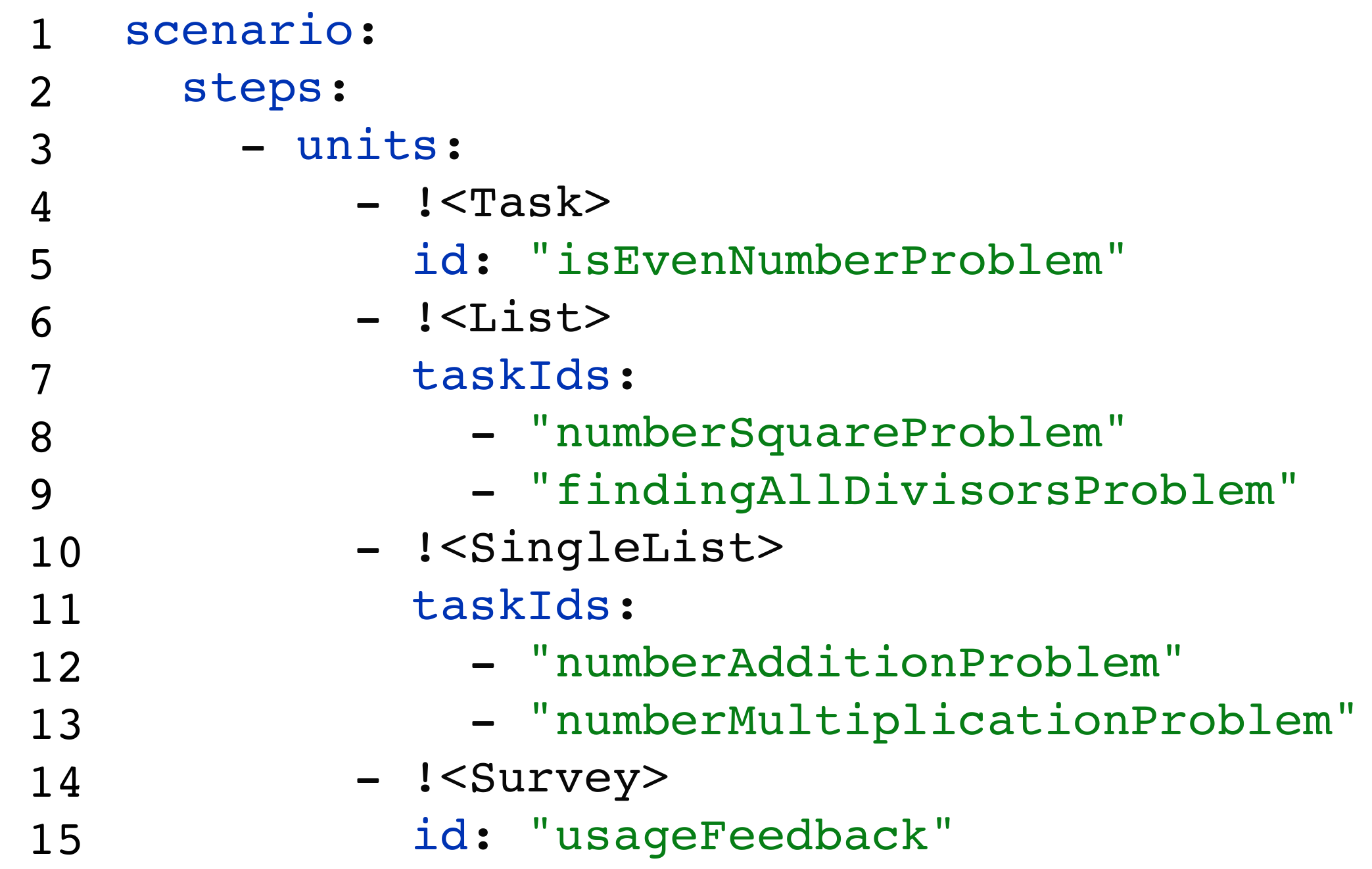}
    \vspace{-0.3cm}
    \caption{Example of a \textit{Scenario Config}.}
    \vspace{-0.3cm}
    \label{fig:scenario}
\end{figure}

\textit{Scenario Config} is an essential part of the plugin responsible for the order in which different actions and tasks are carried out during the education process. This file specifies the order of events, such as solving tasks, completing surveys, displaying information to the student, and others. This sequence will be presented to the student during their learning process, we call this sequence of actions a \textit{scenario}. An example of a simple scenario configuration can be seen in Figure~\ref{fig:scenario}. This scenario contains a sequence of first offering the \textit{isEvenNumberProblem} task, then offering two more tasks in arbitrary order, followed by a choice between two final tasks, of which only one needs to be completed. Finally, the scenario includes showing a survey to the student.

\textit{Task Content Config} specifies the description of the tasks and all files in which the student will solve these tasks. These files are configured relative to the root of the current project in the IDE to prevent any leakage of personal user information. Files are not created if they already exist and are marked in the config as \textit{internal}---this allows collecting files from third-party plugins or educational tools. If a file does not exist, it is created with the content specified in the config or empty if the template was not specified.

\textit{IDE Settings and Inspections Config} contains the IDE settings that should be enabled while implementing the task. Examples of settings that can be changed to study the education process include enabling or disabling specific code quality inspections, enabling or disabling automatic code completion, changing the color theme, etc. This config also controls which code quality inspections should be enabled.

\textit{Survey Config} describes all the surveys that are used in a scenario configuration. Each of these surveys contains a series of different questions, which can be multiple-choice, single-choice, open-ended, etc. A question can be marked as required.

\textit{Activity Data Config} controls how the IDE activity data, such as IDE actions, hotkeys, etc., is processed. This configuration can specify activities that should not be tracked, and also determine the frequency at which they should be tracked.

\textit{Code Tracking Config} consists of rules for tracking file snapshots, such as defining the frequency of code capture or the way in which code snapshots should be taken, such as recording the code in full, recording only function names, and others. Thus, it is possible to collect the full contents of files after every new keystroke or implement a more coarse-grained approach like only collecting function names when the file is saved, tailored to any particular analysis.

\textit{Research Metadata Config} defines the information about a specific study that is being conducted, including its title, description, and other meta-information. This config also contains information about third-party files and logs that are relevant to the specific experiment and should be collected and sent to the server.

\textbf{Privacy policy}. Ensuring the confidentiality of user data during data collection is a crucial aspect of this work. To achieve this, the \toolName plugin has been designed to request students' consent\footnote{The consent link should be specified in the configuration files in accordance with the institutional policy carrying out data collection.} to collect their personal data, such as name, email, and their interactions in the IDE during the study. As it is not desirable to collect unnecessary data, the plugin only reads code from specific files defined in the \textit{Task Content Config} and \textit{Research Metadata Config}. In addition, the data is not collected until students explicitly agree to the user consent on the starting page of the plugin. The link to the specific consent and conditions, as well as their description, are also configurable.

\subsection{Server}
Employing the \toolName server in sync with the plugin offers a convenient solution for data collection. 
The server operates on the {Ktor framework}~\cite{ktor}, providing easy configuration for diverse user needs. The server not only provides the required information for user identification, but also accepts and processes the data from the plugin.
The plugin already comes with a Docker image for the server, so all the user has to do is deploy it to any cloud provider.  
To make the plugin work correctly with a deployed server, the server address must be specified in the plugin configuration file. 

After receiving the data, the server sorts it according to the categories shown in Figure~\ref{fig:collected_data}, and then stores the data in the corresponding database table. The server preserves the raw data as a safeguard against potential processing errors. Regular database backups are also performed as an additional security measure.

\subsection{Visualization Dashboard}

A visualization dashboard is a website where researchers can upload the data collected with \toolName and get basic statistics.\footnote{Visualization dashboard: \url{https://vis-server.labs.jb.gg/}.} The data is not stored on any server, which means that all dashboards are rebuilt every time the page is refreshed. This was intentionally done for privacy reasons. However, the source code of the dashboard is open source and available,\footnote{Visualization dashboard source code: \url{https://github.com/JetBrains-Research/koala-data-vis}.} it comes with the plugin, can be deployed anywhere and connected to any available database if needed. We used the \textit{Streamlit framework}, which provides an easy and fast way to build interactive visualizations. 

\subsection{\textit{ProgSnap2} Converter}

In order to increase the flexibility of \toolName in terms of analysis, we implemented a converter of the gathered data into the \textit{ProgSnap2}~\cite{progsnap2} format. \textit{ProgSnap2} is widely recognized in the research community, and there are many existing tools that work with this format~\cite{CloudCoder, codeworkout, xie2023developing}. Since this format has a predetermined structure, the data collected by different researchers and from different experiments can be combined for a joint analysis. Converting the data collected with \toolName into this format allows to easily share student activity in the IDE and code snapshots with the research community.

\section{Case Study}
\label{section:evaluation}

To evaluate and to highlight the usefulness and the versatility of \toolName, we collected a dataset of authentic student solutions and used it to perform a simple preliminary analysis. In this section, we describe the study, the resulting dataset, and provide an example of an insight that can be derived from the data. The full collected dataset can be found in supplementary materials~\cite{supplementary}.

\textbf{Tasks}. For this case study, we used the recently introduced \textit{in-IDE learning} format~\cite{birillo2024bridging}, in which the students follow a programming course fully within the IDE. We used in-IDE courses provided by the \textit{JetBrains Academy} plugin~\cite{jetbrains-academy-plugin}, which had to be installed along with the \toolName plugin. The \toolName configuration contained relative paths to the task files from the courses, ensuring the correct work of our tool with the \textit{JetBrains Academy} plugin. This highlights how \toolName can be used to track and analyze the education process conducted in other established environments. Specifically, the following courses were used for the experiment:
\begin{itemize}
    \item \textbf{Kotlin Onboarding: Introduction}~\cite{course_onboarding}.
    This course is one of the most popular available courses in \textit{JetBrains Academy}~\cite{jetbrains-academy-marketplace}, tailored for Kotlin beginners and teaching the basic principles of the language. Gathering data from this course can be useful for analyzing how students behave when learning the very basics of a new programming language, in which they have no experience.
    
    \item \textbf{Introduction to IDE Code Refactoring in Kotlin}~\cite{course_refactoring}. This course offers an introduction to effective refactorings and the corresponding IDE features. It enables students to identify refactoring opportunities in the code and address them using automatic IDE refactorings to enhance code structure, readability, and maintenance. 
   The data collected from this course can be used for future research about how novice programmers employ refactorings.
\end{itemize}

\textbf{Study setup}. The \textit{Task Content Config} contained all files from the courses in which the student has to solve practical tasks. All settings in the \textit{IDE Settings Config} were set to \textit{Default}, since the goal was to collect as much data as possible about how students were learning, without restricting them from using the IDE tools. As for the code snapshots, we collected them at the most granular setting possible---collecting the entire files after each keystroke. 

\textbf{Participants}. To conduct this study, we invited 28 students from two universities---16 first-year Bachelor students and 12 third-year Bachelor students. The \textit{Kotlin Onboarding: Introduction} course was completed by all 28 students. The \textit{Introduction to IDE Code Refactoring in Kotlin} course was completed by 16 first-year students.

\begin{figure}[t]
    \centering
    \includegraphics[width=0.9\linewidth]{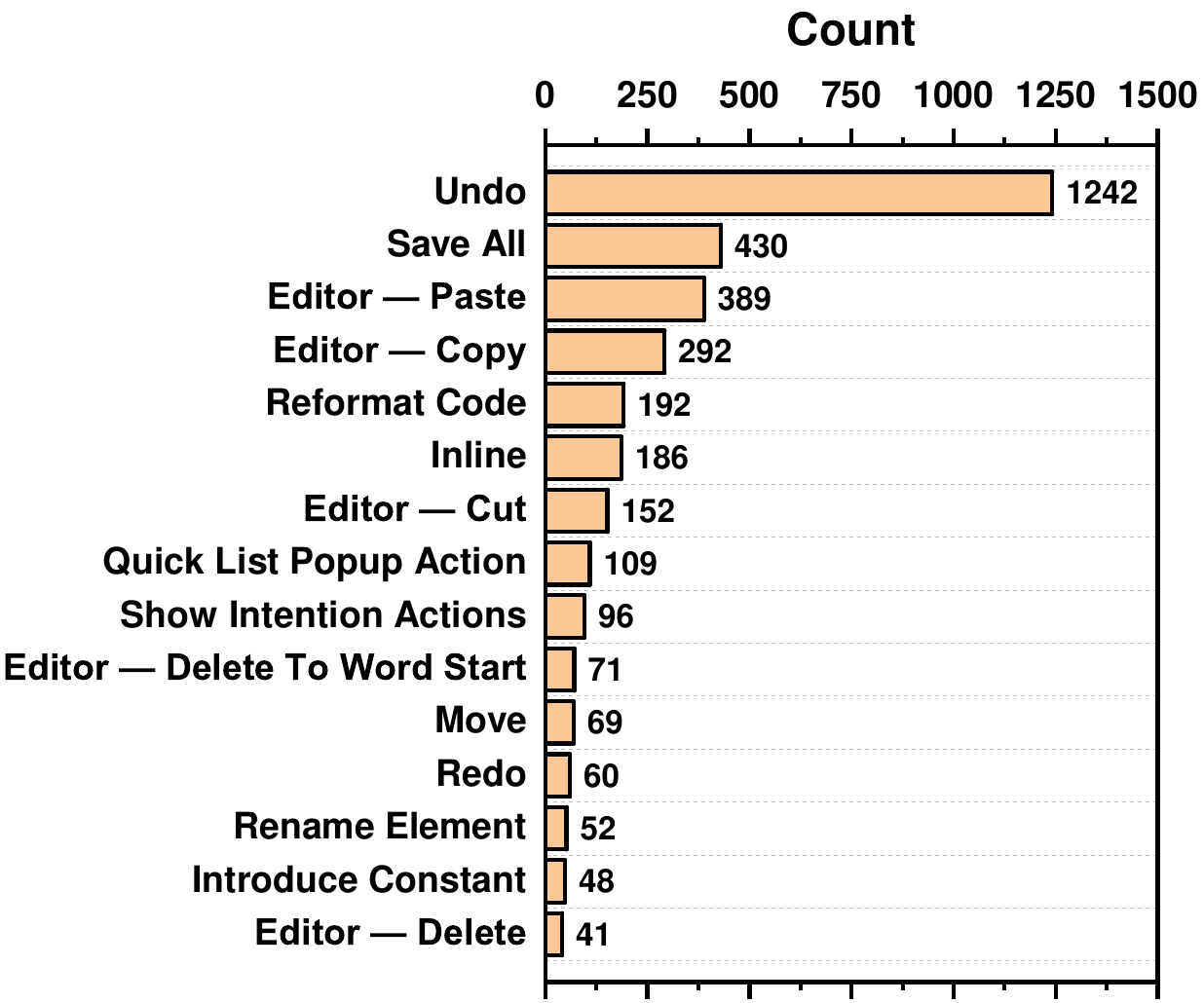}
    \vspace{-0.3cm}
    \caption{Top-15 most used hotkeys when solving tasks in the \textit{Introduction to IDE Code Refactoring in Kotlin} course.}
    \vspace{-0.3cm}
    \label{fig:top_chart}
\end{figure}

\textbf{Collected data}. In total, more than 585 thousand activities in the IDE were collected, among them 94 thousand IDE actions, 3.4 thousand code executions and debugging interactions, and 21 thousand usages of hotkeys. This resulted in over 127 thousand code snapshots, allowing to trace the complete history of writing code in the IDE for each student in each task. Taking the same first-year students that solved tasks in both courses, this data can be used to see how the usage of more advanced features changes as the student gains initial experience.

As an example of possible data analysis, let us consider the newly collected data about using hotkeys. In Figure~\ref{fig:top_chart}, you can see the 15 most used hotkeys when solving tasks in the \textit{Introduction to IDE Code Refactoring in Kotlin} course. We can see what refactorings students conducted with hotkeys as opposed to using IDE toolbars and tool windows, with the most popular ones being \textit{Reformat}, \textit{Inline}, \textit{Move}, and \textit{Introduce constant}. Also, we can compare these hotkeys with more generic ones, such as \textit{Undo}, \textit{Copy}, and \textit{Paste}. 

\textbf{Output format}. The data is saved in the \textit{ProgSnap2} format together with the initial raw collected data. All data is distributed without students' personal information, such as names and emails, and can be used for research purposes. Upon acceptance, we will share the full dataset with the research community, along with the source code of \toolName.

\textbf{Possible applications}. In this paper, we focused on the tool and did not have the space for a deep analysis, thus demonstrating only one possible use of the collected dataset. However, the collected data is rich and can be used for different purposes by other researchers. For example, the analysis of the data in the \textit{Introduction to IDE Code Refactoring in Kotlin} course could show students' misconceptions in advanced refactorings, similar to existing work for basic refactorings~\cite{oliveira2023student}. As for the \textit{Kotlin Onboarding: Introduction} course, it is project-based, and thus its data can be used to design a personalized help system~\cite{roest2024next}.

\section{Conclusion \& Future Work}
\label{section:Conclusion}

In this paper, we presented \toolName, a new tool for collecting code snapshots and IDE activity logs when solving programming tasks in JetBrains IDEs. The \toolName plugin can be easily configured and installed to collect data for various types of research, particularly for in-IDE learning. The plugin is set up using YAML configuration files, which control different aspects of its behavior, allowing for detailed customization.
We also provide an interactive dashboard for the preliminary analysis of the collected data and getting initial insights.
Finally, we implemented a converter that transforms the raw data into the ProgSnap2 format, allowing it to be analyzed by existing tools and combined with other data.

Using \toolName, we carried out an experiment, in which we collected data from 28 students solving tasks in two different in-IDE courses. The resulting data includes more than 585 thousand individual actions and 127 thousand code snapshots. We highlighted how this data and \toolName can be useful for future researchers. In addition, we already used \toolName in a previous study, where it was used to collect not only the problem-solving data, but also additional information about the usage of a developed personalized help system for one of the JetBrains IDEs. 

In future work, we plan to improve the usability of the tool and conduct numerous studies with its help. Specifically, while YAML files allow making the tool highly configurable and versatile, and we made them as intuitive and easy to use as possible, it would be more convenient to have a graphical interface to control the settings. We plan to introduce \toolName into several university curricula and compile a large-scale dataset of students' fine-grained IDE interactions that can be of use to the research community.

\bibliographystyle{ACM-Reference-Format}
\balance
\bibliography{ref}

\end{document}